\documentclass[PRB,twocolumn]{revtex4}
\usepackage{}
\usepackage{epsfig}
\usepackage{graphicx}
\usepackage{amsfonts,amssymb}
\usepackage{amsmath}
\usepackage{color}
\usepackage {times}
\definecolor{refcolor}{rgb}{1.0,0.0,0.0}
\usepackage{lipsum}
\usepackage[colorlinks,citecolor=blue,linkcolor=blue,anchorcolor=blue,filecolor=blue,urlcolor=blue]{hyperref}

\newcommand{\be}{\begin{equation}}
\newcommand{\ee}{\end{equation}}   
\newcommand{\bea}{\begin{eqnarray}}
\newcommand{\eea}{\end{eqnarray}}
\newcommand{\ba}{\begin{array}}
\newcommand{\ea}{\end{array}}

\renewcommand{\k}{{\bf k}}

\begin{document}

\title{Weyl semimetallic state with antiferromagnetic order in Rashba-Hubbard model}

\author{Aastha Jain, Garima Goyal, and Dheeraj Kumar Singh}
 
\affiliation{Department of Physics and Materials Science,
Thapar Institute of Engineering and Technology, Patiala 147004, Punjab, India}

\date{\today}


\date{\today}
\begin{abstract}
We study the phase diagram of Rashba-Hubbard model by employing the Hartree-Fock meanfield theory, and thereby establish the existence of an antiferromagnetically ordered Weyl semimetallic state with in-plane magnetic moments. This phase is found to be sandwiched in between the antiferromagnetic insulator and Rashba metal in the interaction vs spin-orbit coupling phase diagram. The antiferromagnetically-ordered topological semimetallic state exists in the presence of combined time-reversal and inversion symmetry though individually both are broken. The  study of the static magnetic susceptibility indicates the robustness of the antiferromagnetic order within a realistic range of interaction and spin-orbit coupling parameters. In addition to the edge states associated with the Weyl points, we also investigate the spin-resolved quasiparticle interference, which provides important insight into the possible spin texture of the bands especially in the vicinity of Weyl points.   
\end{abstract}

\maketitle

\section{Introduction}   
Different phases are classified according to the different symmetries that may spontaneously be broken at low temperature because of a variety of interactions inherent in material systems, and that can lead to the intriguing diverse phases and associated features~\cite{anderson, klitzing, bernevig, fu, hsieh, konig, xia, zhang, qi}. One such important interaction, which has attracted considerable attention in recent times, and which can profoundly affect the properties of various phases as well the nature of phase transition, is the spin-orbit coupling (SOC), \textit{i.e.}, the entanglement of spin and orbit degrees of freedom ~\cite{winkler, abrikosov, orlenko}. Investigation of proximity effects in hetero-junctions of superconductors and magnetic systems suggests that the SOC may play a critical role in shaping of magnetic, transport, and other exciting properties~\cite{gingrich, niu, linder, ptok}. Signature of non-trivial topological states accompanied with Majorana quasiparticles~\cite{alicea, beenakker, elliot1,fu1, tewari} have been obtained in the systems with SOC.

In the two-dimensional systems, especially at the interface of heterostructures, absence of the inversion symmetry can generate the spin-orbit coupling (SOC), which is referred to as Rashba SOC~\cite{rashba}. The Rashba SOC depends linearly on the crystal momentum $\k$ and lifts the spin degeneracy of energy bands, and therefore can significantly influence the electronic and transport properties giving rise to a wide variety of fascinating properties~\cite{ bychkov, rashba2, rashba3}. Moreover, by controlling the layer thickness of the heterostructure or by the application of an external electric field, the strength of Rashba SOC can be controlled, making such systems more suitable for several potential technological applications~\cite{nitta, schapers, bandyopadhyay, jungwirth, manchon, chappert, knill}. This is unlike the centrosymmetric systems with transition metals as indispensable constituents, where the SOC results from the $LS$ coupling in the $d$ and $f$ orbitals.

The SOC in a correlated-electron system introduces quantum frustration, which, in addition to affecting the nature of Mott transition, can stabilize plethora of exotic states of matter. These states of matter can include axion insulator, spin-orbit coupled Mott insulator, topological semimetals and insulators etc~\cite{krempa, dario, jacob, manchon, joel, kubo}. Although in non-centrosymmetric system, despite significant progress made in understanding different aspects of the Rashba SOC, relatively less attention has been paid to the consequences of its interplay with the  electronic correlation. In a few steps taken in this direction, the nature of metal-to-insulator transition (MIT) and the phase diagram in the Rashba-Hubbard model were studied by using a variety of methods which include Hartree-Fock (HF) approximation~\cite{kennedy}, variational-Monte Carlo (VMC)~\cite{kubo1}, cluster dynamical mean-field theory (CDMFT)~\cite{xzhang} etc.  
 
 Sine-deformed mean-field theory (SDMFT) based on the HF approximation predicts an incommensurate spin-density wave (SDW) states even when the SOC may be vanishingly small and spiral orders for a larger SOC~\cite{chisa_hota}. Other HF-based theories suggest that a metallic state with antiferromagnetic (AFM) order for moderate electronic correlation, AFM insulator for strong electronic correlation, and a striped-magnetic order when both SOC and electronic correlations being large, are stabilized~\cite{kennedy}. The existence of AFM order for smaller SOC is also supported by CDMFT~\cite{xzhang} as well as plaquette-based  investigations~\cite{valentina}. VMC studies point out the existence of Weyl semimetallic (WSM) state without AFM order for stronger SOC and electronic correlation~\cite{kubo1}. Among various phases, thus, found, the metallic AFM state appears to be of considerable interest particularly given its pseudogap like behavior. The current study attempts to provide a detailed investigation of this phase, most importantly, revealing its topological nature.
 
 Our finding suggests that for a reasonable range of SOC and on-site Coulomb interaction, a significant portion of the SOC vs interaction phase diagram is occupied by the metallic and insulating AFM ordered state with in-plane magnetic moment. The transition from the Rashba metal to AFM insulator does not happen directly upon increasing the electronic correlation parameter $U$. Instead, a WSM state with AFM order is stabilized in between the aforementioned two phases. The associated Weyl points (WPs) shift from the high-symmetry points $\Gamma$ and M in the paramagnetic Rashba metallic state to the points along the high-symmetry directions given by $\pm k_x \mp k_y = \pi$ and $k_x$ = -$k_y$ in the WSM-AFM state. Their respective winding numbers are 1 and -1. We also examine the spin-resolved quasiparticle interference (QPI) patterns which is capable of revealing the nature of electronic state including the spin texture in the vicinity of the WPs. 
 
The layout of this paper is as follows. Section II describes the Rashba-Hubbard Hamiltonian, mean-field Hamiltonian, static magnetic susceptibility calculation in the unordered state, and spin-resolved QPI. In section III, the robustness of the AFM ordering, the phase diagram in the  $U-\lambda$ space, the necessary conditions for the WSM state, linearized dispersions near WPs, winding numbers and edge states associated with the WPs, and QPI patterns in the WSM-AFM state are presented. In  section IV, we provide a brief discussion in reference to previous works. Finally, we conclude in section V.

\section{Model and Method}
\subsection{Model}
We consider the one-orbital Rashba-Hubbard Hamiltonian defined on a  square lattice and given by 
\begin{eqnarray} 
    \mathcal{H} = H_t + H_U + H_R,
\end{eqnarray}  \label{1}
where
\begin{eqnarray}
     H_t = -t\sum_{<{\bf i},{\bf j}>}\sum_\sigma (d_{{\bf i}\sigma}^{\dagger}  d^{}_{{\bf j} \sigma} + h.c.) 
\end{eqnarray}
is the kinetic energy term representing the delocalization energy gain due to the nearest-neighbor hopping. $t$ is the hopping amplitude and $d_{{\bf i} \sigma}^{\dagger}$ ($d^{}_{{\bf i} \sigma}$) is the operator creating (annihilating) an electron at site ${\bf i}$ with spin $\sigma$. The unit of interaction parameters and energy henceforth is set to be $t$.

The second term
\begin{equation}
H_U = U \sum_{\bf i} \hat{n}_{{\bf i}\uparrow} \hat{n}_{ {\bf i} \downarrow} \\ 
 \end{equation}
accounts for the on-site Coulomb repulsion between the electrons of opposite spins, where $\hat{n}_{{\bf i} \sigma} = d^\dagger_{{\bf i} \sigma}d^{}_{{\bf i} \sigma}$ denotes the number operator.
 
 The third term $H_R$ represents the Rashba SOC which is defined as
 \begin{equation}
   H_R =  \lambda \sum_{{\bf i},\sigma,\sigma^\prime} [ \textit{i}(d^\dagger_{{\bf i},\sigma}\sigma_{\sigma\sigma^\prime}^x d^{}_{{\bf
   i}+\hat{y},\sigma^\prime} \\
   - d^\dagger_{{\bf i},\sigma}\sigma_{\sigma\sigma^\prime}^y d^{}_{{\bf i}+\hat{x},\sigma^\prime})+ h.c.],
 \end{equation}
\noindent where $\lambda$ is the strength of SOC and $\sigma^{i}$ is one of the Pauli matrices. 

After Fourier transformation, Eq. (2) in the momentum space is given by
\begin{equation}
  H_t({\bf k}) = \sum_{{\bf k},\sigma} \epsilon_{\bf k} d_{{\bf k}\sigma}^\dagger d^{}_{{\bf k} \sigma}
\end{equation}
with $\epsilon_{\bf k} = -2t(\cos k_x + \cos k_y)$, whereas the Rashba term [Eq. (4)] takes the following matrix form
\begin{equation}
    H_R({\bf k}) = \sum_{{\bf k},\sigma,\sigma'}d_{{\bf k} \sigma}^\dagger[\beta_{\bf k} \sigma_{\sigma \sigma^\prime}^x   -   \gamma_{\bf k} \sigma_{\sigma \sigma^\prime}^y]  d_{{\bf k} \sigma'}
\end{equation}

with $\beta_{\bf k} = 2\lambda  \sin k_y$ and $\gamma_{\bf k} =  2\lambda \sin k_x$. 

\subsection{Hartree-Fock mean-field theory}
The Hubbard term $H_U$ being quartic in terms of electron-field operators, has been treated in the absence of Rasbha SOC via variety of techniques based on the mean-field theoretic approaches, perturbation techniques, DMFT~\cite{georges}, QMC~\cite{vekic}, VMC~\cite{kubo2} etc. Here, we use the static mean-field approach based on the Hartree-Fock approximation in order to decouple the interaction Hamiltonian as our focus is mainly on the low-temperature phases. The bilinear term in the electron-field operator, thus, obtained is
\begin{equation}
{H}_{im} 
=  {- \frac{U}{2} \sum_{{\bf i} \sigma} \psi^{\dagger}_{\bf i}(\bf { \sigma}\cdot{\bf m}_{\bf i}) \psi^{}_{\bf i}},
\vspace{-3mm}
\end{equation}
where $\psi^{\dagger}_{\bf i} = (d^{\dagger}_{{\bf i} \uparrow}, d^{\dagger}_{{\bf i} \downarrow}) $. The $j$-th component of magnetic moment at the site ${\bf i}$ is ${m}^{j}_{\bf i} = \frac{1}{2} \langle \psi^{\dagger}_{\bf i}{\sigma}^{j} \psi^{}_{\bf i} \rangle$ with ${\bf m}_{\bf i}$ is the magnetic moment. ${\sigma}^{j}$ is $j$-th component of Pauli matrices. When the decoupled interaction term is incorporated, the Hamiltonian in the basis $({ d^{}_{\bf k \uparrow}, d^{}_{\bf k \downarrow}, d^{}_{\bf k+Q \uparrow}, d^{}_{\bf k+Q \downarrow}})^{T}$ with wavevector ${\bf Q} = (\pi,\pi)$ becomes
\begin{eqnarray}   
  \mathcal{H}_{HF}  =  \sum_{{\bf k}}  (\psi_{{\bf k}\sigma}^\dagger   \psi_{{\bf k+Q} \sigma}^\dagger)
\begin{pmatrix}
  
   \hat{h}_{\bf k}         &    \sigma \cdot {\bf \Delta}               \\
{\bf \Delta}^\dagger \cdot \sigma  &     \hat{h}_{\bf k+Q}  

\end{pmatrix}
\begin{pmatrix}
    \psi_{{\bf k}\sigma} \\
    \psi_{{\bf k+Q}\sigma}
\end{pmatrix},
\end{eqnarray} 

where

\begin{equation*}
    \hat{h}_{\bf k} = 
    \begin{pmatrix}
         \epsilon_{\bf k}      &  \beta_{\bf k} - i\gamma_{\bf k}  \\
        \beta_{\bf k} + i\gamma_{\bf k}  &  \epsilon_{\bf k}  \\
    \end{pmatrix}
\end{equation*}
and 

\begin{equation*}
  {\bf \sigma} \cdot {\bf \Delta} = 
    \begin{pmatrix}
        \Delta_z         &     \Delta_x - i\Delta_y  \\
        \Delta_x + i\Delta_y  &  -\Delta_z  \\
    \end{pmatrix}.
\end{equation*}
Here, $ {\bf \Delta}$ is the exchange field given by $2{\bf \Delta} = U{\bf m}$, where ${\bf m}=m_x\hat{x}+ m_y \hat{y} + m_z \hat{z}$ represents the magnetic-moment vector in one of the sublattices. The direction of the magnetic-moment vector will be opposite in the other sublattice. The two-fold degenerate eigenvalues of the Hamiltonian [Eq. (8)] can be shown to be
\begin{eqnarray}
    E_k &=& \pm \Bigg[\pm 2\bigg(\sqrt{\epsilon_k^2 \beta_k^2 + \epsilon_k^2 \gamma_k^2 + \beta_k^2 \Delta_y^2 + \gamma_k^2\Delta_x^2 - 2\beta_k \gamma_k \Delta_x \Delta_y}\bigg)\nonumber \\ &+&   (\epsilon_k^2 + \beta_k^2 + \gamma_k^2 + \Delta_x^2 + \Delta_y^2)\Bigg]^{\frac{1}{2}}.
\end{eqnarray}
The components of the magnetic moments are given in terms of the eigenvectors of the Hamiltonian as follows:
 \begin{eqnarray} 
m_z &=&\sum_{{\bf k},l}(\phi^{*}_{ {\bf k}\uparrow l} \phi^{}_{{\bf k+Q}\uparrow l} - \phi^{*}_{ {\bf k}\downarrow l} \phi^{}_{{\bf k+Q}\downarrow l}) f(E_{\k l})  \nonumber\\
m_x &=& \sum_{{\bf k},l}(\phi^{*}_{{\bf k}\uparrow l} \phi^{}_{{\bf k+Q}\downarrow l} + \phi^{}_{{\bf k}\uparrow l} \phi^{*}_{{\bf k+Q}\downarrow l})  f(E_{\k l}) \nonumber\\
m_y &=& \sum_{{\bf k},l}(-i\phi^{*}_{ {\bf k}\uparrow l} \phi^{}_{{\bf k+Q} \downarrow l} + i \phi^{*}_{ {\bf k}\uparrow l} \phi^{}_{{\bf k+Q}\downarrow l}) f(E_{\k l}) , \nonumber\\
\end{eqnarray}
where $l$ is the band index and $f(E)$ is the Fermi-Dirac distribution function. 
\subsection{Magnetic susceptibility}
In the previous subsection, we only discussed the Hamiltonian for the commensurate AFM order. Whether a commensurate or incommensurate AFM ordered state with in-plane or out-of-plane magnetic moments can be stabilized, it can be ascertained by examining the static susceptibility given by the 4$\times$4 matrix
\begin{equation}
\hat{\chi}^{0}({\bf q})= 
\left(
\begin{array}{llll}
\chi^{0}_{\uparrow \uparrow \uparrow \uparrow}({\bf q})& \chi^{0}_{\uparrow \downarrow \uparrow \uparrow}({\bf q})& \chi^{0}_{\uparrow \uparrow \downarrow \uparrow}({\bf q})
&\chi^{0}_{\uparrow \downarrow \downarrow \uparrow}({\bf q})\\
\chi^{0}_{\uparrow \uparrow \uparrow \downarrow}({\bf q})& \chi^{0}_{\uparrow \downarrow \uparrow \downarrow}({\bf q})& \chi^{0}_{\uparrow \uparrow \downarrow \downarrow}({\bf q})
&\chi^{0}_{\uparrow \downarrow \downarrow \downarrow}({\bf q})\\
\chi^{0}_{\downarrow \uparrow \uparrow \uparrow}({\bf q})& \chi^{0}_{\downarrow \downarrow \uparrow \uparrow}({\bf q})& \chi^{0}_{\downarrow \uparrow \downarrow \uparrow}({\bf q})
&\chi^{0}_{\downarrow \downarrow \downarrow \uparrow}({\bf q})\\
\chi^{0}_{\downarrow \uparrow \uparrow \downarrow}({\bf q})& \chi^{0}_{\downarrow \downarrow \uparrow \downarrow}({\bf q})& \chi^{0}_{\downarrow \uparrow \downarrow \downarrow}({\bf q})
&\chi^{0}_{\downarrow \downarrow \downarrow \downarrow}({\bf q})
\end{array}
\right) \; 
\end{equation} 
in the presence of SOC~\cite{greco,yanase}. A matrix element of the spin susceptibility $\chi^{0}_{\sigma_1 \sigma_2 \sigma_3 \sigma_4} ({\bf q})$ is defined in terms of the two Green's functions in the unordered state as follows:
\begin{eqnarray}
&&
\chi^{0}_{\sigma_1 \sigma_2 \sigma_3 \sigma_4} ({\bf q})
= 
\sum_{ {\bf k}} G^{0}_{\sigma_1 \sigma_4}({\bf k}) 
G^{0}_{\sigma_3 \sigma_2}( {\bf k}+ {\bf q}),  \qquad
\label{chi0}
\end{eqnarray}
\noindent where  
\begin{equation}
G^{0} ( {\bf k} ) = 
\left[- \hat{h} ({\bf k} ) \right]^{-1} 
\label{2x2Green}.
\end{equation}
In the presence of SOC, the directional spin susceptibilities in different directions may be different. Therefore, they can indicate whether the magnetic ordering with in-plane or out-of-plane magnetic moment is preferred. The spin susceptibilities along three orthogonal directions are $\chi^{0}_{xx}$, $\chi^{0}_{yy}$, and $\chi^{0}_{zz}$ given by
\begin{eqnarray}
\chi^{0}_{xx}({\bf q}) &=& \chi^{0}_{\uparrow \downarrow \uparrow \downarrow}({\bf q})+ \chi^{0}_{\uparrow \downarrow \downarrow \uparrow}({\bf q})+\chi^{0}_{\downarrow \uparrow \uparrow \downarrow}({\bf q})+\chi^{0}_{\downarrow \uparrow \downarrow \uparrow}({\bf q}) \nonumber\\
\chi^{0}_{yy}({\bf q}) &=& -\chi^{0}_{\uparrow \downarrow \uparrow \downarrow}({\bf q})+ \chi^{0}_{\uparrow \downarrow \downarrow \uparrow}({\bf q})+\chi^{0}_{\downarrow \uparrow \uparrow \downarrow}({\bf q})-\chi^{0}_{\downarrow \uparrow \downarrow \uparrow}({\bf q}) \nonumber\\
\chi^{0}_{zz}({\bf q}) &=& \chi^{0}_{\uparrow \uparrow \uparrow \uparrow}({\bf q})- \chi^{0}_{\uparrow \uparrow \downarrow \downarrow}({\bf q})-\chi^{0}_{\downarrow \downarrow \uparrow \uparrow}({\bf q})+\chi^{0}_{\downarrow \downarrow \downarrow \downarrow}({\bf q}).
\end{eqnarray}
The largest peak for the spin susceptibility $\chi^{0}_{xx}({\bf q})$ or $\chi^{0}_{zz}({\bf q})$ occurring at a wavevector ${\bf q} = {\bf Q} $ implies an inherent instability against magnetic ordering with the wavevector ${\bf Q}$. 

\subsection{DOS modulation due to single impurity}
Quasiparticle interference is a powerful technique to study the low-energy quasiparticle excitations in a system~\cite{jennifer1}. The elastic scattering of quasiparticles by an impurity atom generates interference patterns, which may provide important insight especially into the electronic band structure in the vicinity of the Fermi level. The modulation in the local density of states (DOS) corresponding to the interference patterns can be calculated with the help of Green's function.

The change induced in Green's function due to a single impurity with $\delta$-potential can be studied by using $t$-matrix approximation~\cite{guo, zhang1}. This change is  
\begin{equation}
    \centering
      \delta \hat{G} ({\bf k, k^{\prime}}, \omega) = \hat{G}^{0} ({\bf k}, \omega) \hat{T} (\omega) \hat{G}^{0} ({\bf k^{\prime}}, \omega) 
\end{equation}
    where $\hat{G}^{0} ({\bf k}, \omega) = (\omega \hat{I}  - \hat{\mathcal{H}}_{HF})^{-1}$ is the free particle Green's function in AFM ordered state.
$\hat{T}$ matrix of scattering by an impurity atom is defined in terms of the Green's function given by  
\begin{equation}
    \centering
       \hat{T} (\omega) = (\hat{{I}} -\hat{V}^{imp} \mathcal{\hat{G}}^{0} (\omega))^{-1} \hat{V}^{imp},
\end{equation}
where the Green's function summed over all momenta in the Brillouin zone is
\begin{equation}
    \centering
       \mathcal{\hat{G}}^{0} (\omega) = \frac{1}{N} \sum_{\bf k} \hat{G}^{0} ({\bf k}, \omega)
\end{equation}
  and
\begin{equation}
    \centering
      \hat{V}^{imp}_i = {\hat{V}}_{i} \otimes \hat{\sigma}_{i}, \nonumber
\end{equation}  
where
\begin{equation}
    \centering
      \hat{V}_{i} = V_{oi}
    \begin{pmatrix}
     1   &   1 \\
     1   &   1
    \end{pmatrix}.
\end{equation} 
$\hat{V}_{i}^{imp}$ denotes the $4 \times 4$ scattering matrix due to an impurity atom, which is written in terms of Pauli matrices $\sigma_i (i=1,2,3)$ representing various spin-resolved channels of impurity scattering. $i = 0$ corresponds to scattering due to non-magnetic impurity. We also investigate the behavior of interference patterns generated when the tip of the probe is able to resolve the spin state of the quasiparticle. The spin-resolving tip $(\hat{V}_{tip})$ can be described in terms of Pauli matrices as
\begin{equation}
    \centering
      \hat{V}^{tip}_i = {\hat{V}}_{i} \otimes \hat{\sigma}_{i} \nonumber
\end{equation}  
where
\begin{equation}
    \centering
      \hat{V}_{i} = V_{ot} 
    \begin{pmatrix}
     1   &   1 \\
     1  &   1
    \end{pmatrix}.
\end{equation} 
Now, the change $\delta \rho_{ij} ({\bf k}, \omega)$ recorded by a spin-resolving tip in the DOS due to the scattering by an impurity atom can be given by~\cite{guo} 
\begin{eqnarray}
    \centering
       \delta \rho_{ij}({\bf q}, \omega) &=& -\frac{1}{2 \pi} \sum_{\bf k} {\rm Im} [{\rm Tr} \hat{V}^{tip}_i \hat{G}^{0}({\bf k},\omega) \hat{T}(\omega) \nonumber\\ &\times&  \hat{G}^{0}({\bf k + q},\omega)].
\end{eqnarray}
\section{Results}
\subsection{Magnetic-ordering instability}
\begin{figure}
    \centering
    \includegraphics[scale=1, width= 8.8cm]{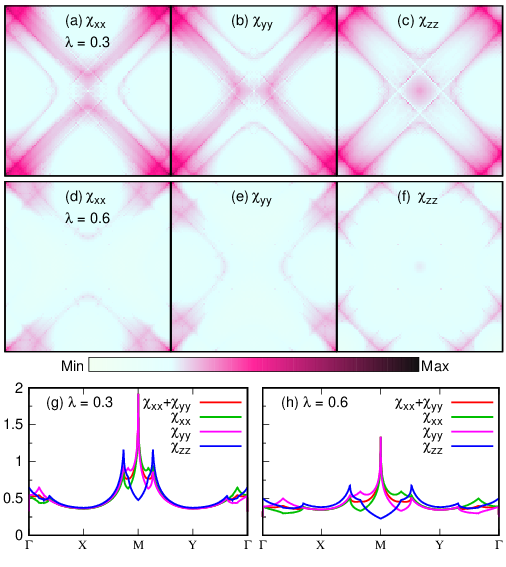}
    \caption{Various components of the static spin susceptibility in the whole Brillouin zone $(a$-$f)$ and also along the high-symmetry directions $(g$-$h)$ showing peaks for $\lambda = 0.3$ and $0.6$. }
    \label{sus}
\end{figure}

\begin{figure}
    \centering
  \includegraphics[scale=10.8, width=8.8cm]{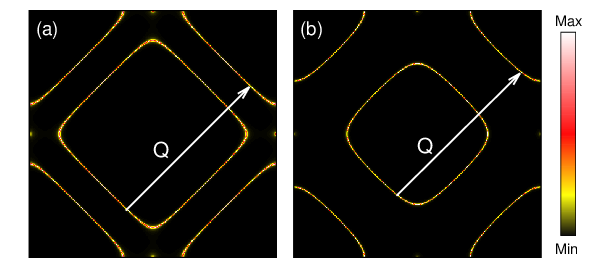}
    \caption{Fermi-surfaces with one of the prominent nesting vectors ${\bf Q} = (\pi,\pi)$ at half-filling for (a)  $\lambda = 0.4$ and (b) $\lambda = 0.8$, where the range of both $k_x$ and $k_y$ is $[-\pi,\pi]$ .}
    \label{fs}
\end{figure}

We begin by examining the magnetic ordering  instability in the Rashba-Hubbard model. Fig.~\ref{sus} shows the static magnetic susceptibilities in different directions for the unordered state. Because of the SOC, $SU(2)$ rotational symmetry is absent, therefore, the in-plane component $\chi^{0}_{xx}(\bf q)$ and the out-of-plane component $\chi^{0}_{zz}(\bf q)$ are different. Moreover, $\chi^{0}_{xx}(\bf q)$ and $\chi^{0}_{yy}(\bf q)$ are also different along the high-symmetry directions such as $\Gamma$-X, $\Gamma$-Y, M-Y etc. We find that both $\chi^{0}_{xx}(\bf q)$ or $\chi^{0}_{yy}(\bf q)$ show stronger divergences at ($\pi, \pi$) as compared to $\chi^{0}_{zz}(\bf q)$, when $\lambda$ is small, implying that the magnetic moments will preferably be oriented in the $x-y$ plane. Secondly, upon increasing $\lambda$, peak position for $\chi^{0}_{zz}(\bf q)$ shifts away from $(\pi, \pi)$ to an incommensurate wavevector. There is no similar shift in the peak position for the in-plane component of the magnetic susceptibility. The latter is a robust feature in a realistic range for the SOC parameter, which is a result of nesting between the Fermi pockets around $\Gamma$ and M points with nesting vector ($\pi, \pi$) [Fig.~\ref{fs}]. In the large SOC limit, \textit {$t/\lambda \rightarrow$ 0}, we find that $\chi_{xx}$ and $\chi_{yy}$ are peaked at $(0, \pi)$ and $(\pi, 0)$, respectively, whereas $\chi_{zz}$ at (0, 0) instead. Interestingly, all the three peaks have the same height indicating the same strength of instability against ferromagnetic or stripe order. The former set of peaks arise from the interpocket scattering between the pockets around $(0, 0)$ and ($\pi, \pi$). The latter peak originates from the intra pocket scattering. It may be noted that, in the $t/\lambda  \rightarrow 0$, the tiny pockets formed around $(0, 0)$ and ($\pi, \pi$) becomes Weyl points~\cite{kubo2}.

\begin{figure}
    \centering
    \includegraphics[scale=1.0, width=8.8cm]{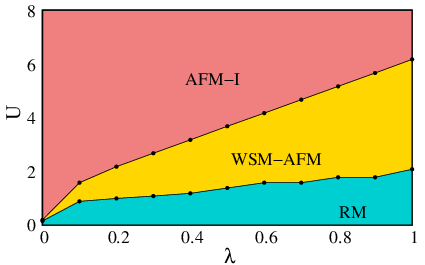}
    \caption{Phase-diagram showing various phases for the correlation strengths $0 \le U \le 8$ and Rashba SOC strength $ 0 \le \lambda \le 1$. Three different phases including AFM-I (antiferromagnetic insulator), WSM-AFM (Weyl semimetallic antiferromagnet), and RM (Rashbha metal).}
    \label{phase}
\end{figure}
Upon inclusion of electron-electron interaction within the random-phase approximation (RPA), the susceptibilities are expected to  diverge at the peak positions for a critical value of $U$ and a given $\lambda$, indicating a singularity in the free energy, and hence the instability of the system against a magnetic order. Our findings, thus, suggest that the commensurate AFM order with in-plane magnetic moments will be stabilized.    
\subsection{WSM-AFM state}
Having examined the robustness of the AFM state with in-plane magnetic moment, we now explore the phase diagram in the $U- \lambda$ parameter space, by using the self-consistent mean-field theoretic approach discussed above. Since the in-plane RPA-level magnetic susceptibility is expected to show divergence at the wavevector ($\pi, \pi$), the initial value of out-of-plane magnetic moment, \textit{i.e.}, $m_z$ is set to be zero without any loss of generality. Only two types of phases are expected to occur, one with self-consistently obtained magnetic order, and another without magnetic order. Throughout the calculations, the inverse temperature parameter 
($\beta = 1/kT)$ is fixed to be 1000 with $k$ as the Boltzmann constant, which corresponds to $\sim5$K, if $t \sim$ 0.5eV.  

\begin{figure} 
    \centering
    \includegraphics[scale=1.0, width=9.0cm]{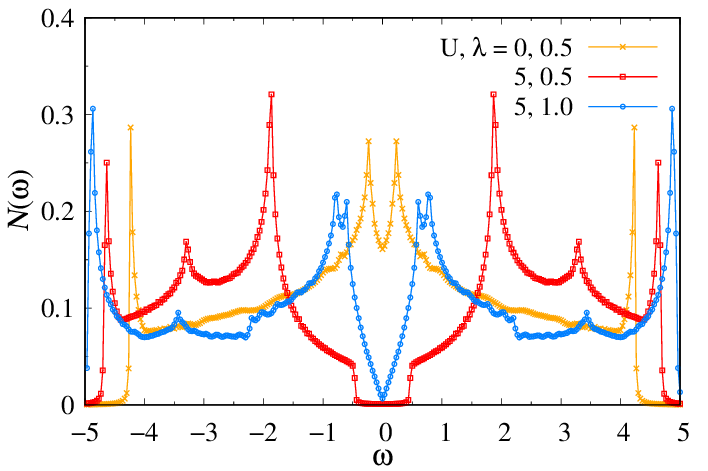}
    \caption{DOS corresponding to the three different phases, RM, AFM-I, and WSM-AFM. For $U = 0$ and $\lambda = 0.5$, a significant DOS is obtained at Fermi level  characteristic of a metallic system. In case of $U = 5$ and $\lambda = 0.5$, the DOS is gapped at the Fermi level indicating the insulating behavior of the system. Upon increasing SOC further so that $\lambda = 1$, the DOS is zero at the Fermi level, whereas it does not vanish in the immediate vicinity, indicating a semimetallic state.}
    \label{dos}
\end{figure}

\begin{figure} 

    \centering
    \vspace{-30mm}
    \includegraphics[scale= 1.5, width=10.5cm]{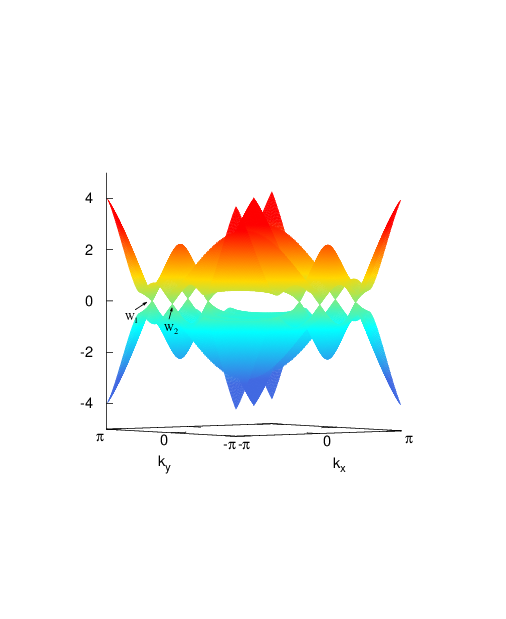}
    \vspace{-30mm}
    \caption{ Electronic dispersion in the WSM-AFM shown for the entire Brillouin zone.}
    \label{disp1}
\end{figure}

Fig.~\ref{phase} shows the phase diagram, which consists of phases with magnetic order as well as the Rashba metal (RM) phase without magnetic order. There are two types of magnetically ordered states. One is the AFM insulator (AFM-I) occurring for higher $U$ and other is the topological WSM state with magnetic order (WSM-AFM) existing for relatively lower value of $U$ to be discussed at length later, as it is the main focus of the current work. In an earlier work~\cite{kennedy}, a pseudogap-like phase with AFM order was reported based on the finding of a dip in the DOS [Fig. ~\ref{dos}], which is similar to the one observed in the so-called pseudogap phases found in the hole-doped cuprates~\cite{kanigel,singh}. However, the dispersion plotted along various high-symmetry directions $(0,0) \xrightarrow{}(\pi,\pi) \xrightarrow{} (0,\pi) \xrightarrow{} (-\pi/2,\pi/2)$ provides a hint for the topological nature of this magnetically-ordered state as there exist two pairs of linear non-degenerate band crossings at the Fermi level giving rise to two pairs of symmetry protected Weyl points (WP) [Fig.~\ref{disp} \&\ref{disp1}]. 

The Hamiltonian ${\mathcal H}_{HF}$ lacks both the time-reversal invariance ($\mathcal{T}$) and inversion ($\mathcal{P}$) symmetries, when considered independently. In order to see it, we rewrite the mean-field Hamiltonian in the sublattice basis formed because of AFM ordering as 
\bea
{\mathcal H}_{HF} &=& \epsilon_{\bf k} \sigma_0 \otimes \tau_1 +  \Delta_x \sigma_1 \otimes \tau_0  +\Delta_y\sigma_2 \otimes \tau_0 \nonumber\\ &+& 2\lambda \sin k_y \sigma_1 \otimes \tau_1 -2\lambda 
\sin k_x \sigma_2 \otimes \tau_1. 
\eea
Here, $\sigma_i$s and $\tau_j$s are the Pauli matrices used for the spin and sublattice degrees of freedom. It may be noted that only a few of the five matrices formed by $\sigma_i$s and $\tau_j$s in the Hamiltonian anticommute, \textit{i. e.}, they don't form Dirac algebra. 

All the terms in the Hamiltonian do not commute even for $\Delta_x = \Delta_y = 0$, \textit{i. e.}, when the AFM order is absent. In that case, along $\pm k_x \mp k_y = \pi$, the first term contribution vanishes, the remaining two terms anticommute. Then, the Hamiltonian has both time-reversal ($\mathcal{T} = i\sigma_2 K$) and inversion ($\mathcal{P} = \tau_1$) symmetries, which leads to the two-fold degenerate bands along (0, $\pi$) $\rightarrow$ (-$\pi$/2, $\pi/2$) and Dirac points (DPs) at (0, $\pi$) and ($\pi$, 0). This is mainly the consequence of Brillouin zone folding. In the absence of magnetic ordering, the folding is not required, then the degeneracy disappears and the DPs change into the WPs. The degerancy of the bands is lifted and the DPs are split into WPs if the system develops non-zero magnetic moment. One of the WPs is shifted away from (0, $\pi$) along $(0, \pi) \rightarrow (-\pi/2, \pi/2)$. Another band degeneracy occurs along ($\pi, 0$) $\rightarrow$ (0, $\pi$), \textit{i.e.}, along $k_x + k_y = \pi$, which persists even in the AFM ordered state. This is a consequence of the fact that the first term vanishes while the rest anticommute with each other. 

With magnetic order, both the time-reversal and inversion symmetries are broken. It is not difficult to see that the second and third terms break the time-reversal symmetry while the fourth and fifth terms are responsible for the absence of inversion symmetry. Although, the Hamiltonian is invariant under the combined operation defined by $\mathcal{T} \mathcal{P}$, \textit{i.e.}, $\mathcal{T} \mathcal{P} (\mathcal{H}_{HF}(\k)) (\mathcal{T} \mathcal{P})^{-1} \rightarrow \mathcal{H}_{HF}(\k)$. Unlike the system considered here without inversion symmetry, earlier works show that stable DPs can be obtained in the nonsymmorphic system with second-neighbor SOC in the absence of time-reversal and inversion symmetries while both combinedly remain intact~\cite{wang1}. 

The bands crossing at the Fermi level in the WSM-AFM state disappears with a rise in the value of $U$. The boundary between the AFM-I and WSM-AFM state is determined by the condition $|\Delta| \leq 2\lambda$, which should be satisfied by the magnetic exchange coupling $\Delta$. On lowering $U$ further, there is a phase transition from WSM-AFM to RM with the magnetic moment vanishing to zero in the self-consistent scheme. This happens because the spin susceptbility does not diverge at ${\bf Q}$ for a given set of parameters, which includes on-site Coulomb interaction. In the following, we obtain the condition to be satisfied by the magnetic order parameter to determine if the AFM ordered state is Weyl semimetallic or not.
First of all, we note that the self-consistently obtained in-plane magnetization has the same magnitude along both directions, \textit{i.e.}, $\Delta_x = \Delta_y = \Delta$. If the condition $|\Delta| \le 2\lambda$ is satisfied then one pair of WPs ($W_1$) is located along $k_x - k_y = \pi$ and $-k_x + k_y = \pi$ directions whereas the other pair ($W_2$) along $k_x = - k_y$ as seen in Fig.~\ref{disp}. Note that the word ``pair'' does not necessarily mean that they have opposite winding number as to be discussed later, it is used here only to indicate their location in the Brillouin zone.
\subsection{WPs along $ \pm k_x \mp k_y = \pi$ }
\begin{figure} 
    \centering
    \includegraphics[scale= 1.0, width=9.0cm]{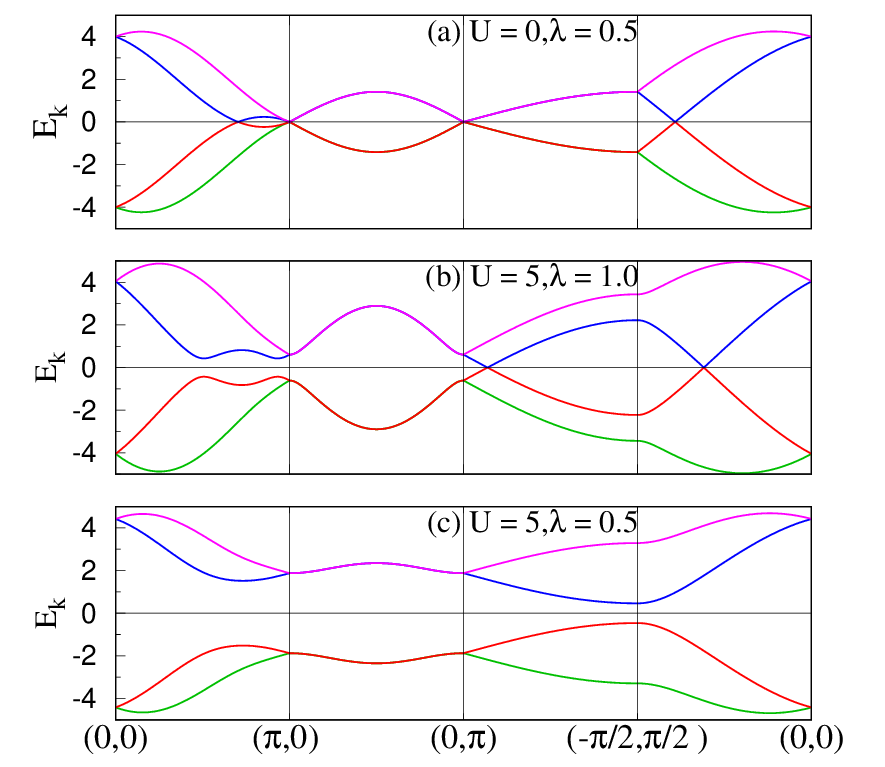}
    \caption{Electronic dispersions are plotted for different sets of $U$ and $\lambda$ values. They correspond to the three phases  (a) RM, (b) WSM-AFM, and (c) AFM-I.}
    \label{disp}
    \end{figure}

For the pair of Weyl points $W_1$ located along $k_x - k_y = \pi$ and $-k_x + k_y = \pi$ [Fig.~\ref{disp}], $\epsilon_{\bf k} = 0$  and  $\beta_{\bf k} = -\gamma_{\bf k}$. Therefore, the eigenvalues [Eq. (9)] of the Hamiltonian matrix in the magnetically ordered state reduce to  
\begin{equation}
    E_{1 \bf k } = \pm\sqrt{2(\beta_{\bf k} - \Delta)^2}.
    \label{eqn.9}
\end{equation}
As the dispersion near WPs (say $(k_{x_o},k_{y_o})$) is linear, Taylor-series expansion around these points by replacing  $k_x^\prime \xrightarrow{} k_{x_o} + q_x$ and $k_y^\prime  \xrightarrow{} k_{y_o} + q_y$ with small ${\bf q}$ gives 
\begin{eqnarray}
    E_{1{\bf q}} = \pm \sqrt{2} \Big[ 2 \lambda (\sin k_{y_o} + q_y \cos k_{y_o}) - \Delta \Big].
    \end{eqnarray}
The dispersion can be linear and the band crossing will appear at $(k_{x_o},k_{y_o})$ provided $2 \lambda \sin k_{y_o} = \Delta$.  In that case
\begin{equation}
    E_{1 {\bf q}} = \pm c q_y
\end{equation}
where c is a constant term given with $c = 2\sqrt{2}\lambda \Bigg[\sqrt{1 - {\Big(\frac{\Delta}{2\lambda}}\Big)^2}\Bigg]$. Since $|\sin k_{y_o}| \le 1$, a linearized dispersion will be obtained at $k_{y_o} = \sin^{-1} \frac{\Delta}{2\lambda}$ whenever the self-consistently computed exchange field satisfies the condition $|\Delta| \leq 2\lambda$. 

\subsection{WPs along $k_x = -k_y$}
Next, we examine the linear-band crossing for the other pair of WPs $W_2$, which is found along $k_x = -k_y$. Along this direction, $\beta_{\bf k} = -\gamma_{\bf k}$ and $\epsilon_{\bf k} = -4t \cos k_y$. Therefore, the energy eigenvalues reduce to  
\begin{equation}
    E_{2 \bf q} = \pm {\bigg (}\sqrt{(-4t \cos k_y)^2 + 2 \Delta^2} - \sqrt{2} \beta_{\bf k} \bigg)
    \label{eqn.12}.
\end{equation}
Now introducing the Taylor-series expansion around the momenta corresponding to the $W_2 = (k_{x_o}, k_{y_o})$, we obtain the linearized dispersion 
\begin{eqnarray}
     E_{2{\bf q}} = \pm c^{\prime} q_y,
\end{eqnarray}
 where
 \begin{eqnarray}
   c^{\prime} = \Bigg(\frac{16 t^2 b^\prime}{\sqrt{b}} + 2\sqrt{2}\lambda\Bigg) (\sqrt{1- b^{\prime 2}})
 \end{eqnarray}
and
 \begin{equation}
     b = \Big(4t\sqrt{1 - b^{\prime 2}}\Big)^2 +2\Delta^2. 
 \end{equation}
 The linearized dispersion and WPs are possible only when $|b^{\prime}| \le 1$ where   
 \begin{equation}
 b^\prime = \sin k_{y_o} = \sqrt{\frac{8t^2 + \Delta^2}{4\lambda^2 + 8t^2}}.
 \end{equation} 
 This again yields the condition $ |\Delta| \le 2\lambda$ same as in the case of $W_1$ points. The location of $W_2$ can be obtained with the help of Eq.~(29) by calculating $k_{y_o}$.

\subsection{Chern numbers for WPs $W_1$ and $W_2$ } 

\begin{figure}
    \centering
    \includegraphics[scale=0.8, width=8.8cm]{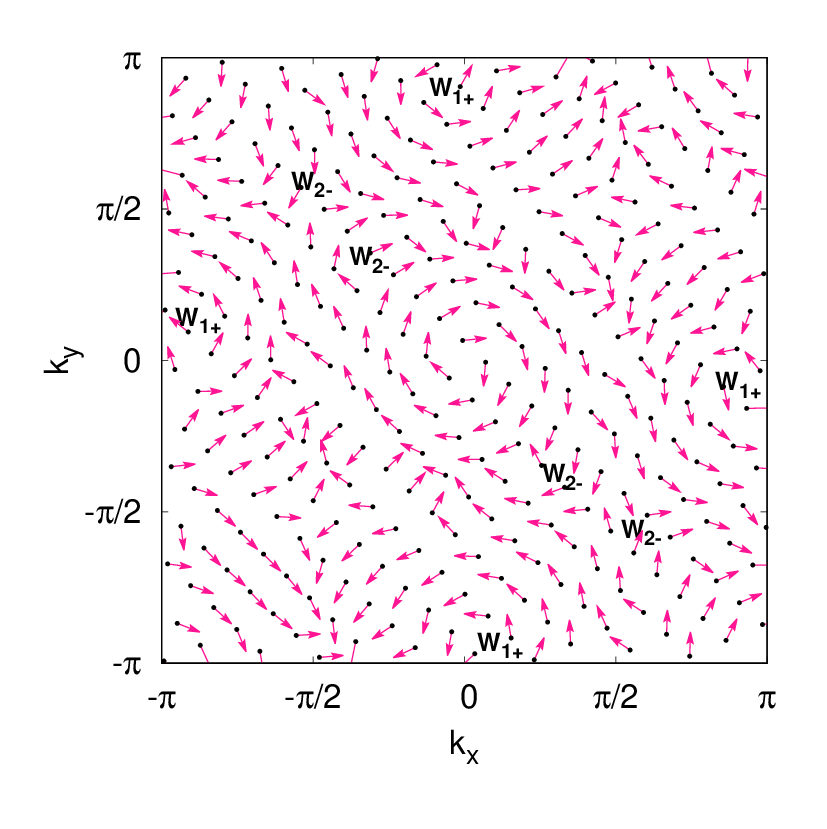}
    \vspace{-1.0cm}
    \caption{ Berry connection $A^{}({\bf k})$ plotted for the whole Brillouin zone. The nature of winding about each WP is denoted by either $W_{+}$ or $W_{-}$, where positive and negative signs indicate counterclockwise and clockwise windings, respectively.}
    \label{WeylP1}
\end{figure}

\begin{figure}
    \centering
    \includegraphics[scale=0.8, width=8.8cm]{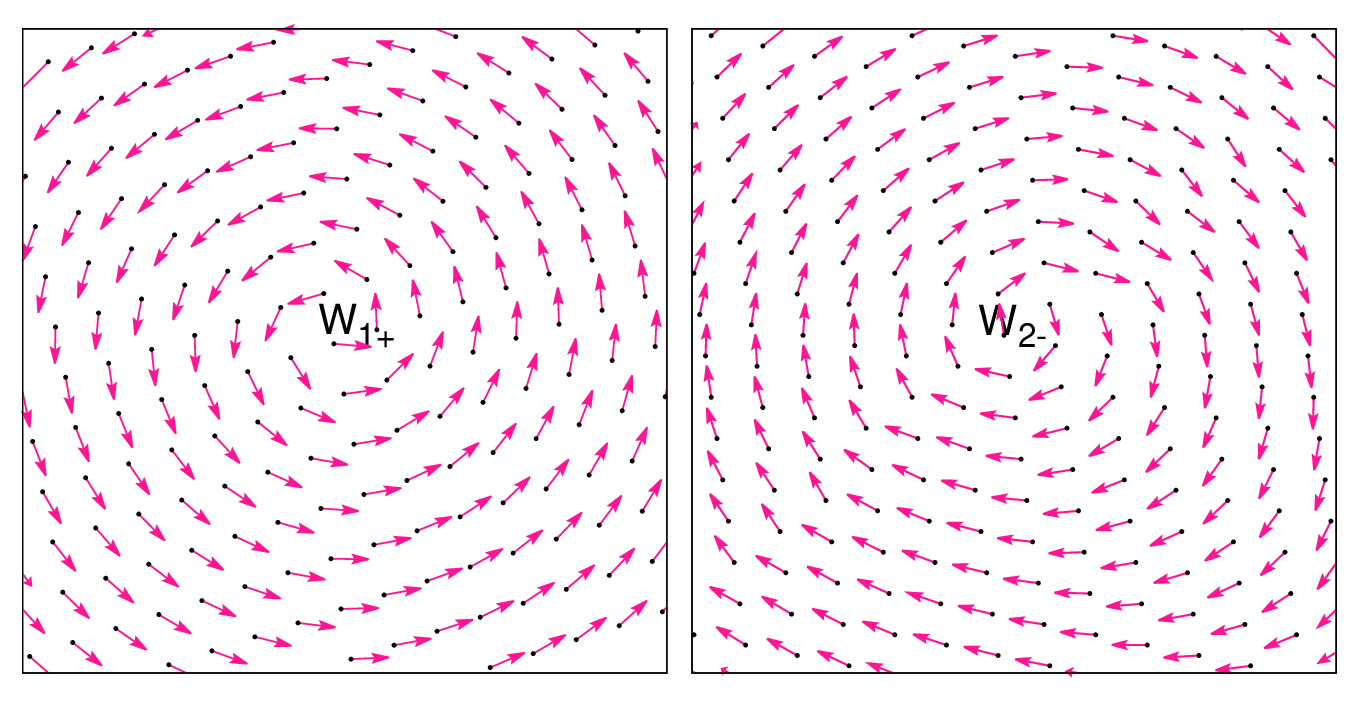}
    \caption{Zoomed in view of the Berry connection about the Weyl points shown in the Fig. 7 as the rotation of the fields are not clear especially in the case of $W_{2-}$.}
    \label{WeylP2}
\end{figure}

\begin{figure}
    \centering
    \includegraphics[scale=0.8, width=8.8cm]{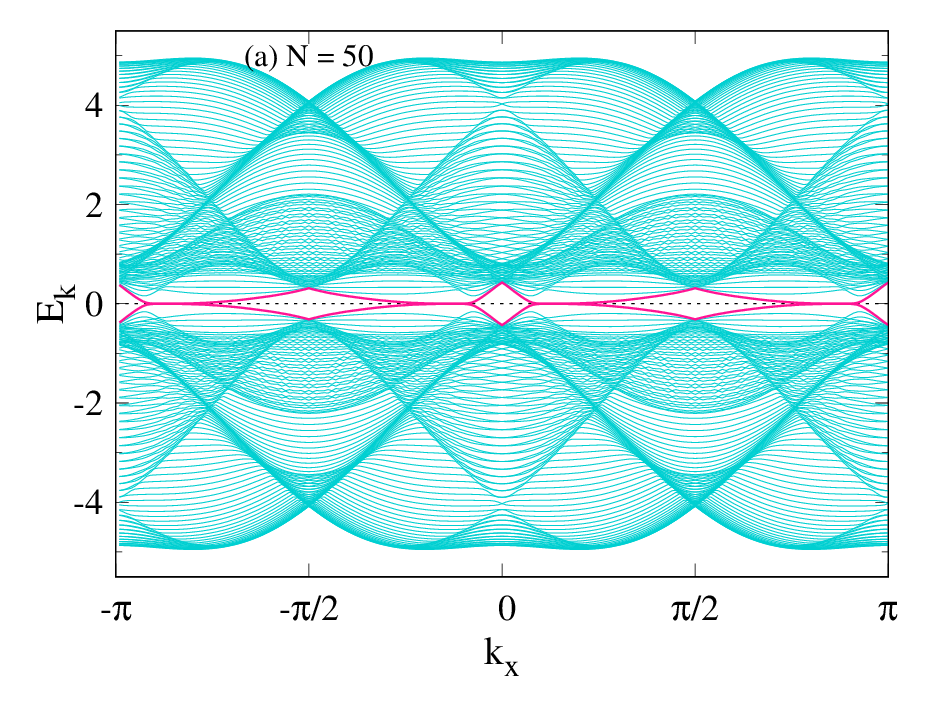}\\
     \includegraphics[scale=0.8, width=8.8cm]{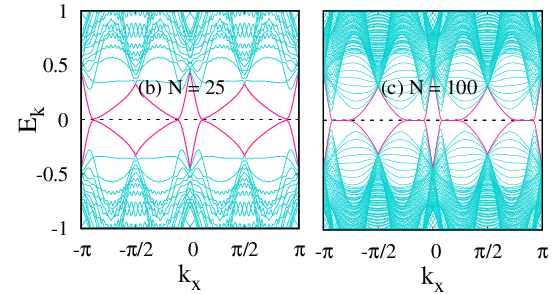}
    \caption{For the set of parameters $U = 5$ and $\lambda = 1$, (a) the  edge-state and bulk band dispersions are plotted for a ribbon of width $N = 50$ extended along $x$-direction and projected onto one dimensional Brillouin zone. The edge states crossing the band gap are colored differently for demarcation. Figure also shows the zoomed in version for different size (b) $N = 25$ and (c) $N = 100$.}
    \label{edgestate}
\end{figure}
In the previous subsections, we focused on the linearized dispersion in the vicinity of WPs $W_1$ and $W_2$. Next, we address the question of winding numbers associated with these WPs. In the unordered state, the WPs occurring at (0, 0) and ($\pi, 0$) have winding numbers 1 and -1, respectively. However, in the magnetic Brillouin zone, the calculation of winding numbers using Eq.~(9) analytically may be a difficult task, therefore, we adopt a numerical approach. The winding number for the pair of WPs $W_1$ and $W_2$ can be obtained by calculating the following line integral~\cite{berry} 
\begin{equation}
w = -\frac{i}{2\pi} \oint {\bf A}^{(n)} ({\bf k}) \cdot d{\bf l} ,
\end{equation}
performed along a closed contour enclosing a WP, where the Berry connection ${\bf A}^{(n)} ({\bf k})$ for the $n$th band is given by
\begin{equation}
 {A}^{(n)}_i ({\bf k}) = \langle u_n(\k)| \frac{ \partial}{\partial k_i}|u_n(\k) \rangle .
\end{equation}
$|u_n(\k) \rangle$ is the eigenvector of the Hamiltonian given by Eq. (8). The calculation by using Eq. (29) yields the winding number $w$ for WPs $W_1$ and $W_2$ along $ \pm k_x \mp k_y = \pi$ and $k_x = -k_y$, 1 and -1, respectively. Fig.~\ref{WeylP1} shows the vector field associated with the Berry connection in the entire Brillouin zone while a zoomed in view for two such points with anticlockwise and clockwise rotations is presented in Fig.~\ref{WeylP2}. 

\subsection{Edge States}
As discussed above, there exist WPs inside and on the boundary of reduced Brillouin zone with winding numbers $\pm 1$. Therefore, existence of edge states along a quasi-one dimensional chain is guaranteed in a way similar to the case of graphene~\cite{wakabayashi, berenvig} or other magnetically ordered systems with Dirac points~\cite{lau, wang1}. These edge states, in the case of topological insulator, are pairs of states propagating in direction opposite to each other. The edge-state  dispersion crosses the Fermi level and connects the valence and conduction bands. In order to explore the edge state in the WSM-AFM state, we consider a ribbon oriented along the $x$-direction consisting of $N$ chains of atoms positioned along $y$ direction such that $k_x$ becomes a good quantum number. Thus, the Hamiltonian $H_{Rbx}$ for the ribbon is $4N \times 4N$ matrix given by

\begin{equation}
    H_{Rbx} ({\bf k}) = 
    \begin{pmatrix}
        H_{1}^+  &  H_2  &  O &  H_3 &  \cdots \\
        H^{\dagger}_2  &  H_{1}^-  &  H_3   &  O &  \cdots \\
        O  &  H^{\dagger}_{3}  &  H_{1}^+  &  H^{}_2  & \cdots \\
        H^{\dagger}_{3}  &  O  &  H^{\dagger}_2  &  H_{1}^-  &  \cdots \\
        \vdots  &  \vdots  &  \vdots  &  \vdots  &  \ddots \\
    \end{pmatrix},
\end{equation}

where 

\begin{equation*}
    H_{1}^\pm = 
    \begin{pmatrix}
        0         &     \pm(\Delta_x - i\Delta_y)  \\
        \pm(\Delta_x + i\Delta_y)  &  0 \\
    \end{pmatrix},
\end{equation*}

\begin{equation*}
    H_2 = 
    \begin{pmatrix}
        2t \cos k_x      &  2i\lambda \sin k_x  \\
        -2i\lambda \sin k_x   &  2t \cos k_x  \\
    \end{pmatrix},
\end{equation*}
and
\begin{equation*}
    H_3 = 
    \begin{pmatrix}
        t  &  i\lambda  \\
        i\lambda  &  t  \\
    \end{pmatrix}.
\end{equation*}
In Fig.~\ref{edgestate}, the bands crossing the Fermi energy and shown in different color are correspond to the edge states. Above and below are the bulk dispersion bands, which are gapped. The edge states cross the Fermi level at four different points, two of the crossings lie very close to $-\pi$ and $\pi$, while the other two lie near $k_x = 0$. We find that the crossing turns into a flat degenerate bands as the number of chains in the ribbon is increased. Here, we have focused on the ribbon oriented along $x$ direction. When the ribbon is oriented along the $y$-direction, same edge-state dispersion is obtained, which is the consequence of the reflection symmetry about the line $k_x$ = $k_y$ in the Brillouin zone. Although we have restricted our calculations to the direction along the primitive translational vectors of the original lattice, it would be interesting to study the edge states if the ribbons are oriented along the primitive translational vector of the reduced Brillouin zone, when the chains consist of sites belonging to a particular sublattice only, \textit{i.e.}, the magnetic moments are aligned along the same direction. 

\subsection{Quasiparticle interference}

\begin{figure}[hb]
    \centering
    \includegraphics[scale=1.0, width=8.8cm]{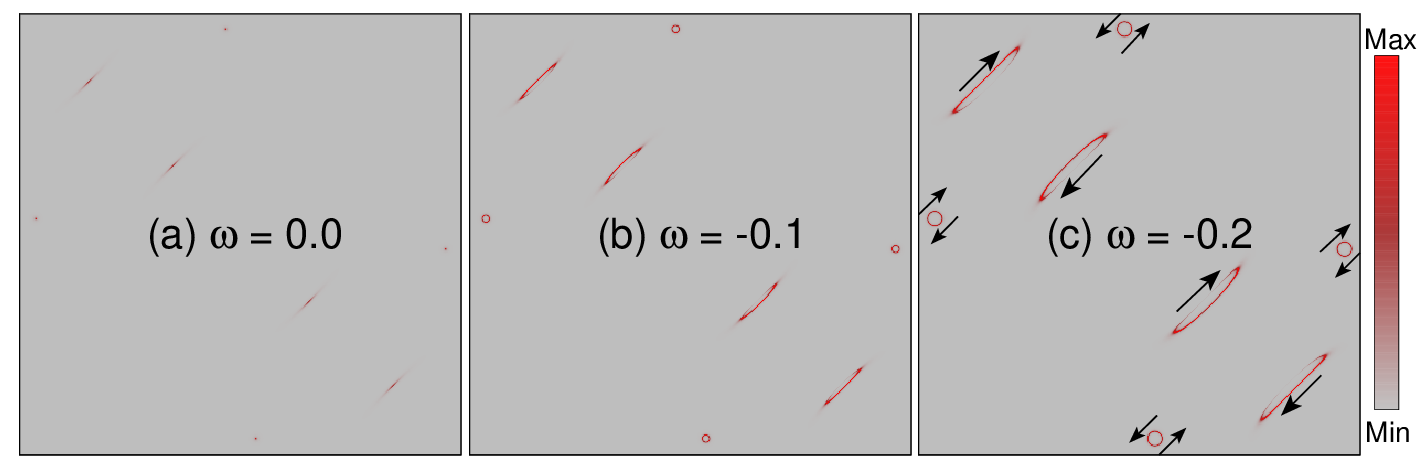}
    \caption{Constant-energy contours (CECs) obtained in the WSM-AFM state for energies (a) $\omega = 0.0$, (b) $\omega = -0.1$, and (c) $\omega = -0.2$. The arrows in (c) indicate orientation of spin along the CECs.}
    \label{cec}
\end{figure}

\begin{figure}[hb]
    \centering
    \includegraphics[scale=1.0, width=8.8cm]{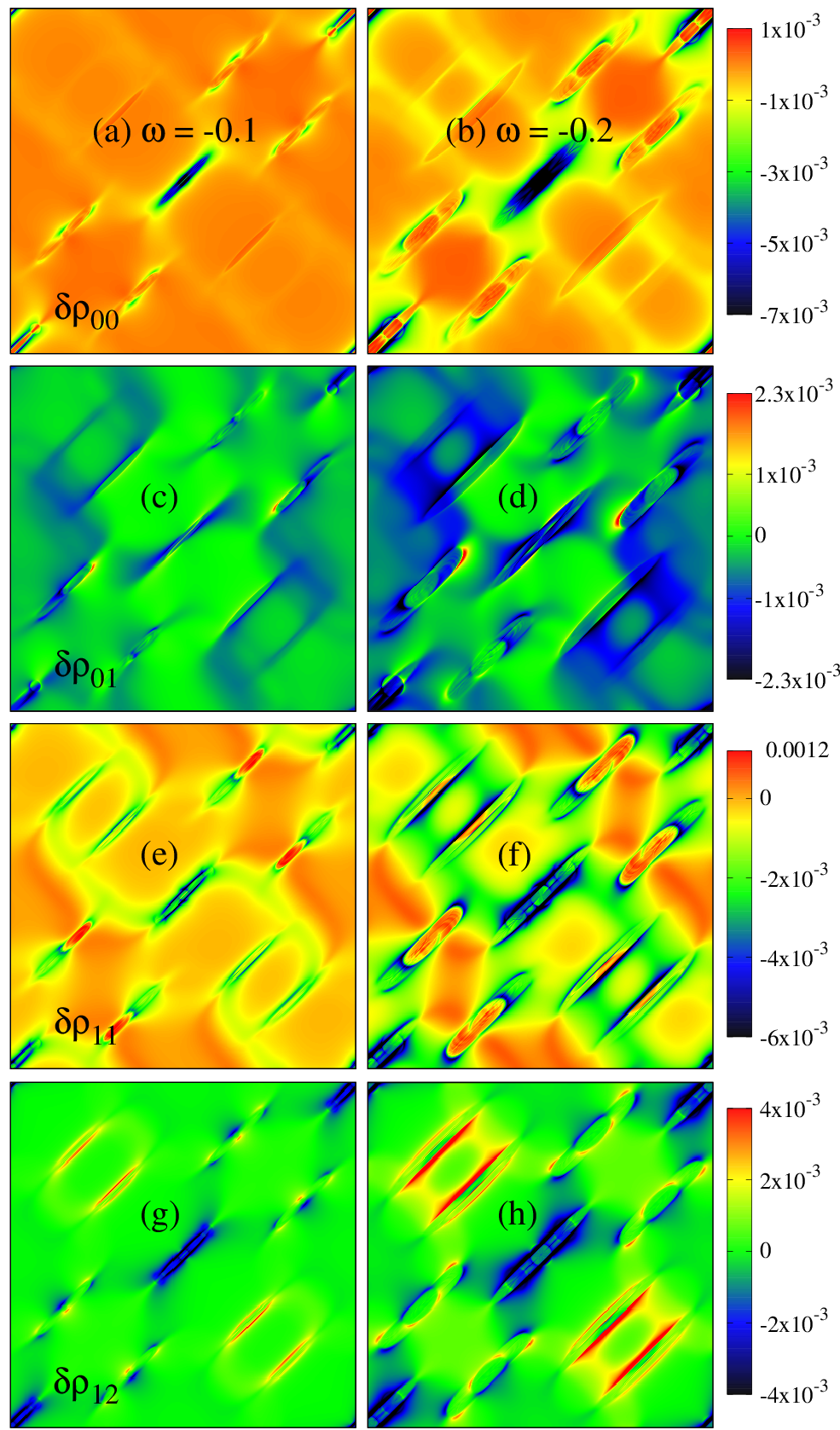}
    \caption{QPI patterns due to non-magnetic impurity $\rho_{00}$ (first row) and magnetic impurity $\rho_{01}$ (second row). Third and fourth row show the QPI patterns $\rho_{11}$ and $\rho_{12}$ detected by spin-state sensitive tip due to a magnetic impurity. The first column correspond to energy $\omega = -0.1$ while those in the second column to $\omega = -0.2$.}
    \label{qp}
\end{figure}

A conclusive signature of Weyl points may be obtained with the help of spin- and angle-resolved photoemission spectroscopy (ARPES)~\cite{prx}. Besides, the scanning-tunneling microscopy (STM), which can detect the impurity generated modulation in the local density of states, \textit{i.e.} QPI, can be another effective tool to confirm Weyl points.

The QPI patterns for a non-topological systems is determined primarily by the spectral-density distribution along the constant-energy contours (CECs) for a given energy as well as by the shape of the contours~\cite{hoffman}. In particular, the patterns are dominated by those scattering vectors, which connect the regions with high spectral density~\cite{roushan}.  However, the situation is contrastingly different for a helical-Fermi liquid in the topological semimetals, where the backscattering by a non-magnetic impurity atom may not allowed, therefore no characteristic response is expected in the patterns. On the other hand, the magnetic impurity generates only a weak response~\cite{guo}. If the tip of the probe can differentiate between the spin state of the quasiparticle, the QPI patterns can provide crucial information about the band structure in the vicinity of Fermi surface as well as the spin texture. Here, in the WSM-AFM, there exist only two species of quasiparticle, either with spin oriented along $\hat{x} + \hat{y}$ or along -$\hat{x}$ - $\hat{y}$, where $\hat{x}$ and $\hat{y}$ are the unit vectors directed along the primitive translational vectors, respectively. The orientation of the  quasiparticle spin is in accordance with the direction of the magnetic moments in the WSM-AFM state. 

 Fig.~\ref{cec} (a) and (b) show CECs for $\omega = -0.1$ and $-0.2$. There are two types of CECs in the magnetic Brillouin zone. One with the circular shape, which has spins pointing along $\hat{x} + \hat{y}$ in one half and along -$\hat{x}$ - $\hat{y}$ in another half. The quasiparticle spin is required to change the direction abruptly at the intersection of a circular CEC and a line running along the zone-diagonal direction and bisecting the CEC. Another set of CECs appear like the cross section of a banana along its length, with highly non-uniform spectral density along. It may be noted that each of these CECs in the magnetic Brillouin zone has the quasiparticle with spin pointing along only one direction. In other words, upon moving along the CECs and completing one cycle, there is no change in the spin direction. Thus, the orientation of quasiparticle spin changes along CECs, which is different from the way it changes along the cross-section of Dirac cone for 
in helical liquid existing on the surface of a strong topological insulator~\cite{guo}.  

Unlike the Dirac cone generated QPI patterns, where the non-magnetic impurities can give rise to only weak and  nonsingular response~\cite{farrell}, the patterns in the WSM-AFM state are not featureless. Fig.~\ref{qp} shows the QPI patterns obtained for $V_{oi} = 0.1$ and $V_{ot} = 0.1$. The main reason is only two possible orientation of spins along the CECs. The intrapocket scattering for the banana-shaped CECs generates a pattern of similar shape at $\Gamma$ as the quasiparticle spin has the same orientation all along. The pattern at $\Gamma$ has also intrapocket contribution due to the circular pockets near the points like $(\pi,0)$, which is weak and unnoticeable because of tiny scattering vectors. The patterns at ($\pm \pi, \mp \pi$) are generated by the interpocket scattering between the banana-shaped CECs belonging to the reduced Brillouin zone and the one outside it, separated by ($\pm \pi, \mp \pi$) in the momentum space. It may be noted that they have same orientation of spin. On the other hand, the interpocket scattering between the two banana-shaped CECs belonging to the reduced Brillouin zone, does not generate any pattern because of opposite spin orientation. The interpocket scattering between the circular and banana-shaped pocket also creates a noticeable pattern because one side of circular pocket has the quasiparticle spin aligned along the spin of quasiparticle associated with the banana-shaped CECs.

Second row shows $\rho_{01}$, the patterns generated by the magnetic impurity with spin along $x$ direction, when the STM is spin insensitive. The pattern generated by the intrapocket scattering from the banana-shaped CECs appears to be weak. More or less similar behavior is observed for the patterns originating from the interpocket scattering. The weak patterns originate because the  quasiparticle spin direction is flipped after impurity scattering. Further, the scattered quasiparticle finds a reduced DOS with spin oriented along positive $x$ direction. Note that this is also true for the quasiparticle having the spin component along the negative $x$ direction. On the other hand, the spin-resolved QPI shows strong patterns, which arise because of the specific orientation of the quasiparticle spin after scattering, and then detected through the spin-sensitive STM probe. $\rho_{11}$ patterns show strong singular features resulting from both the intrapocket and interpocket scatterings. The intrapocket scattering leads to two linear patterns around $\Gamma$ as well as around  $\pm (\pi, \pm \pi)$. The interpocket scattering between the two nearest banana-shaped CECs results into two linear patterns around ($\pm \pi/2$, $\mp \pi/2$). The orientation of the pattern follows directly from the alignment of CECs. QPI patterns because of the interpocket scattering between the circular and banana-shaped CECs are also easily noticeable. These features are more or less repeated for $\rho_{12}$ when the impurity spin is oriented along $y$ direction while the STM probe has spin oriented along $x$ direction. The similarity results from the fact that the quasiparticle spins are not oriented along $x$ or $y$ direction instead along $x$ + $y$ or -$x$ - $y$.

\section{Discussion} 
The search for symmetry-protected two-dimensional topological semimetals similar to graphene is of significant theoretical interest as well as technological applications. These topological semimetals may be protected by the crystalline symmetries and can be destroyed by magnetic ordering as the time-reversal symmetry is broken~\cite{young, matneeva}. However, some of the recent studies indicate that the magnetic order can coexist with topological semimetallic state. For instance, the Dirac points (DPs) were observed in the SDW state of iron pnictides~\cite{richard}. These DPs are protected by the collinearity of the SDW state, inversion symmetry about an iron atom, and invariance under the combined time-reversal and inversion of magnetic moments~\cite{wan}. Similarly, a Dirac-semimetallic (DSM) state has been predicted to exist in a system with nonsymmorphic symmetry, when both ${\mathcal T}$ and ${\mathcal P}$ are broken~\cite{wang1,goyal}. In the current work, we have demonstrated the existence of WSM state with AFM order having checkerboard arrangement of spins within the Rashba-Hubbard model, where the time-reversal and inversion symmetry both are individually absent. However, the combined time-reversal and inversion symmetry is protected even in the absence of nonsymmorphic symmetries.   

In the current work, the phase diagram is obtained at a temperature $T = 0.001$ in the unit of $t$. This particular value of  temperature corresponds to $\sim$ 5 K if $t$ is taken to be $\sim$ 500meV. With rise in the temperature, the magnetic moments can melt away within the static mean-field theory or randomness can be introduced in their orientation if considered within a more sophisticated approaches involving Monte-Carlo simulations based on the auxiliary-field methods~\cite{harun1,harun2}. For this reason, the AFM order makes room for
the RM phase in the phase diagram. A similar phase diagram although without WSM-AFM state has been reported earlier at a very small temperature, when the difference between the free-energy and ground state energy becomes increasingly small~\cite{kennedy}. It is worthwhile to note that when $T \rightarrow 0$ K, the Rashba metallic state will not be found at all. Therefore, the area occupied by the WSM-AFM state will increase upon lowering the temperature. Secondly, with a rise in temperature, as the magnetic moments melt away, $W_1$ will be shifted to the point ($\pi, 0$), whereas $W_2$ will move toward the momentum with $k_{y_o}= \sin^{-1}\big(\frac{8t^2}{4\lambda^2+8t^2}\big)$. A similar consequence on the phase diagram is expected, when the quantum correction to the sublattice magnetization is incorporated, which will reduce the sublattice magnetization~\cite{avinash}. The current approach ignores the spatial and thermal fluctuations in the magnetic moments. It will be of strong interest to see the consequences of such fluctuations on the stability of WSM state, which can be studied most effectively with the exact-diagonalization + Monte Carlo (ED + MC) approach~\cite{harun1,harun2}.

In addition to Hartree-Fock meanfield theory, the $U-\lambda$ phase diagram has been obtained via VMC~\cite{kubo1}. Although, the WSM is found for larger $\lambda$ when the Fermi surface shrinks to a point at the high-symmetry points $(0, 0)$ and $(\pi, \pi)$. Such a Fermi surface structure is unlikely to lead to a robust magnetic ordering, and therefore the corresponding WSM state may exist without any magnetic order. Secondly, the WSM state sandwiched in between the Rashba metal and AFM insulator does not occur for a lower value $\lambda$. Our calculations, on the other hand, demonstrates that if the magnetic moments in the AFM ordered state satisfies the condition $|\Delta| \le 2\lambda$, then, the AFM ordered state can coexist with the WSM state. Therefore, in future studies, it would be of interest to see as to what kind of bandstructure will be supported for the AFM insulating state obtained via VMC method when the condition $|\Delta| \le 2\lambda$ is fulfilled.

Here, we have restricted our effort to half filling with a focus on the previously reported metallic antiferromagnetic state with pseudogap-like features in the density of states~\cite{kennedy}. It would be interesting, in future work, to examine the possible existence of topological states with incommensurate magnetic order stabilized away from half filling. Morever, our calculation shows that the second-neigbhor hopping can also support the Weyl points but not necessarily a Weyl semimetallic state. This is because the pairs of the Weyl points are shifted away from the Fermi level in opposite direction. It may appear that the band filling can be changed in order to force one of the Weyl points to lie at the Fermi level. However, it will be also accompanied by additional bands crossing the Fermi level. In addition, the nature of magnetic order may not necessarily remain to be of the checkerboard-type.
\section{Conclusion} 
In conclusion, we have shown the existence of Weyl semimetallic state with antiferromagnetic order in the Rashba-Hubbard model in a realistic range of interaction and spin-orbit coupling parameters. The Weyl semimetallic state is accompanied with two pair of Weyl points inside the reduced Brillouin zone, where the combination of both inversion and time-reversal symmetries exist when taken together. Although both the symmetries are absent individually because of the Rasbha spin-orbit coupling and magnetic order. In addition, the linear dispersion in the vicinity of Weyl points, winding numbers, and associated edge states dispersion are also studied. Finally, both the spin-sensitive as well as spin-insensitive quasiparticle interference were invstigated, which provided with valuable insight into the nature of quasiparticle spin state in the vicinity of the Weyl points.

\section*{Acknowledgement} 
D.K.S. was supported through DST/NSM/R\&D HPC Applications/2021/14 funded by DST-NSM and start-up research
grant SRG/2020/002144 funded by DST-SERB.

\end{document}